\documentclass[conference,a4paper,dvipsnames]{IEEEtran}

\usepackage[a4paper,left=1.43cm,right=1.43cm,top=1.8cm,bottom=4.43cm]{geometry}

\usepackage{xcolor}
\usepackage{cite}
\usepackage{amssymb}
\usepackage{amsthm,amsmath,array}
\usepackage{graphicx}                  
\usepackage{rotating}
\usepackage{times}
\usepackage{bm}
\usepackage{enumitem}
\usepackage{epsfig,psfrag}
\usepackage[export]{adjustbox}
\usepackage[caption=false,font=footnotesize]{subfig}

\definecolor{darkgreen}{rgb}{0.0, 0.5, 0.0}
\definecolor{magenta_}{rgb}{1.0, 0, 1.0}

\definecolor{darkgrey}{rgb}{0.3, 0.3, 0.3}

\newcommand{\be}{\begin{equation}}
\newcommand{\ee}{\end{equation}}
\newcommand{\ist}{\hspace*{.3mm}}
\newcommand{\rmv}{\hspace*{-.3mm}}

\newcommand{\nn}{\nonumber}

\makeatletter
\newcommand*\bigcdot{\mathpalette\bigcdot@{.5}}
\newcommand*\bigcdot@[2]{\mathbin{\vcenter{\hbox{\scalebox{#2}{$\m@th#1\bullet$}}}}}
\makeatother

\usepackage{bm}

\DeclareMathAlphabet{\mathsfbr}{OT1}{cmss}{m}{n}
\SetMathAlphabet{\mathsfbr}{bold}{OT1}{cmss}{bx}{n}
\DeclareRobustCommand{\msf}[1]{%
  \ifcat\noexpand#1\relax\msfgreek{#1}\else\mathsfbr{#1}\fi
}

\makeatletter
\newcommand{\msfgreek}[1]{\csname s\expandafter\@gobble\string#1\endcsname}
\makeatother

\DeclareFontEncoding{LGR}{}{} 
\DeclareSymbolFont{sfgreek}{LGR}{cmss}{m}{n}
\SetSymbolFont{sfgreek}{bold}{LGR}{cmss}{bx}{n}
\DeclareMathSymbol{\salpha}{\mathord}{sfgreek}{`a}
\DeclareMathSymbol{\sbeta}{\mathord}{sfgreek}{`b}
\DeclareMathSymbol{\sgamma}{\mathord}{sfgreek}{`g}
\DeclareMathSymbol{\sdelta}{\mathord}{sfgreek}{`d}
\DeclareMathSymbol{\sepsilon}{\mathord}{sfgreek}{`e}
\DeclareMathSymbol{\szeta}{\mathord}{sfgreek}{`z}
\DeclareMathSymbol{\seta}{\mathord}{sfgreek}{`h}
\DeclareMathSymbol{\stheta}{\mathord}{sfgreek}{`j}
\DeclareMathSymbol{\siota}{\mathord}{sfgreek}{`i}
\DeclareMathSymbol{\skappa}{\mathord}{sfgreek}{`k}
\DeclareMathSymbol{\slambda}{\mathord}{sfgreek}{`l}
\DeclareMathSymbol{\smu}{\mathord}{sfgreek}{`m}
\DeclareMathSymbol{\snu}{\mathord}{sfgreek}{`n}
\DeclareMathSymbol{\sxi}{\mathord}{sfgreek}{`x}
\DeclareMathSymbol{\somicron}{\mathord}{sfgreek}{`o}
\DeclareMathSymbol{\spi}{\mathord}{sfgreek}{`p}
\DeclareMathSymbol{\srho}{\mathord}{sfgreek}{`r}
\DeclareMathSymbol{\ssigma}{\mathord}{sfgreek}{`s}
\DeclareMathSymbol{\stau}{\mathord}{sfgreek}{`t}
\DeclareMathSymbol{\supsilon}{\mathord}{sfgreek}{`u}
\DeclareMathSymbol{\sphi}{\mathord}{sfgreek}{`f}
\DeclareMathSymbol{\schi}{\mathord}{sfgreek}{`q}
\DeclareMathSymbol{\spsi}{\mathord}{sfgreek}{`y}
\DeclareMathSymbol{\somega}{\mathord}{sfgreek}{`w}

\DeclareMathSymbol{\svarsigma}{\mathord}{sfgreek}{`c}

\DeclareMathSymbol{\sGamma}{\mathalpha}{sfgreek}{`G}
\DeclareMathSymbol{\sDelta}{\mathalpha}{sfgreek}{`D}
\DeclareMathSymbol{\sTheta}{\mathalpha}{sfgreek}{`J}
\DeclareMathSymbol{\sLambda}{\mathalpha}{sfgreek}{`L}
\DeclareMathSymbol{\sXi}{\mathalpha}{sfgreek}{`X}
\DeclareMathSymbol{\sPi}{\mathalpha}{sfgreek}{`P}
\DeclareMathSymbol{\sSigma}{\mathalpha}{sfgreek}{`S}
\DeclareMathSymbol{\sUpsilon}{\mathalpha}{sfgreek}{`U}
\DeclareMathSymbol{\sPhi}{\mathalpha}{sfgreek}{`F}
\DeclareMathSymbol{\sPsi}{\mathalpha}{sfgreek}{`Y}
\DeclareMathSymbol{\sOmega}{\mathalpha}{sfgreek}{`W}

\DeclareRobustCommand{\mcal}[1]{%
  \ifcat\noexpand#1\relax\mathnormal{#1}\else\cal{#1}\fi
}
\DeclareRobustCommand{\BM}[1]{%
  \ifcat\noexpand#1\relax\bm{\boldUppercaseItalicGreek{#1}}\else\bm{#1}\fi
}
\makeatletter
\newcommand{\boldUppercaseItalicGreek}[1]{\csname var\expandafter\@gobble\string#1\endcsname}
\makeatother

\newcommand{\V}[1]{\bm{#1}}
\newcommand{\M}[1]{\BM{#1}}
\newcommand{\Set}[1]{{\mcal{#1}}}

\usepackage{mathtools}

\newcommand{\trans}{^{\mathrm{T}}}

\newcommand{\diag}[1]{\mathrm{diag}}   

\IEEEoverridecommandlockouts
\begin{document}
\allowdisplaybreaks

\title{\huge Multipath-based SLAM using Belief Propagation with Interacting Multiple Dynamic Models}
\author{\IEEEauthorblockN{Erik Leitinger\IEEEauthorrefmark{1}, Stefan Grebien\IEEEauthorrefmark{1}\IEEEauthorrefmark{2}, and Klaus Witrisal\IEEEauthorrefmark{1}\IEEEauthorrefmark{2}}
\\[-2mm]
\small{\IEEEauthorrefmark{1}Graz University of Technology, Graz, Austria (\{erik.leitinger,stefan.grebien,witrisal\}@tugraz.at)} \\ \small{\IEEEauthorrefmark{2}Christian Doppler Laboratory for Location-aware Electronic Systems}
}
\maketitle

\begin{abstract}
  In this paper, we present a Bayesian multipath-based simultaneous localization and mapping (SLAM) algorithm that continuously adapts interacting multiple models (IMM) parameters to describe the mobile agent state dynamics. The time-evolution of the IMM parameters is described by a Markov chain and the parameters are incorporated into the factor graph structure that represents the statistical structure of the SLAM problem. The proposed belief propagation (BP)-based algorithm adapts, in an online manner, to time-varying system models by jointly inferring the model parameters along with the agent and map feature states. The performance of the proposed algorithm is finally evaluating with a simulated scenario. Our numerical simulation results show that the proposed multipath-based SLAM algorithm is able to cope with strongly changing agent state dynamics.
\end{abstract}                            

\section{Introduction}\label{sec:Introduction}
Simultaneous localization and mapping (SLAM) is important in many fields including robotics \cite{Thrun2005}, autonomous driving \cite{BressonTIV2017}, location-aware communication \cite{DiTaranto2014SPM}, and robust indoor localization \cite{GentnerTWC2016, LeitingerTWC2019, WinSPM2018}. Specifically, robustness, i.e. achieving a low probability of localization outage, is still challenging in environments with strong multipath propagation\cite{LeitingerJSAC2015, MendrzikTWC2019, ShahmansooriTWC2018,KulmerTWC2018}. 

\begin{figure}[t]
\centering
\includegraphics[width=0.9\columnwidth]{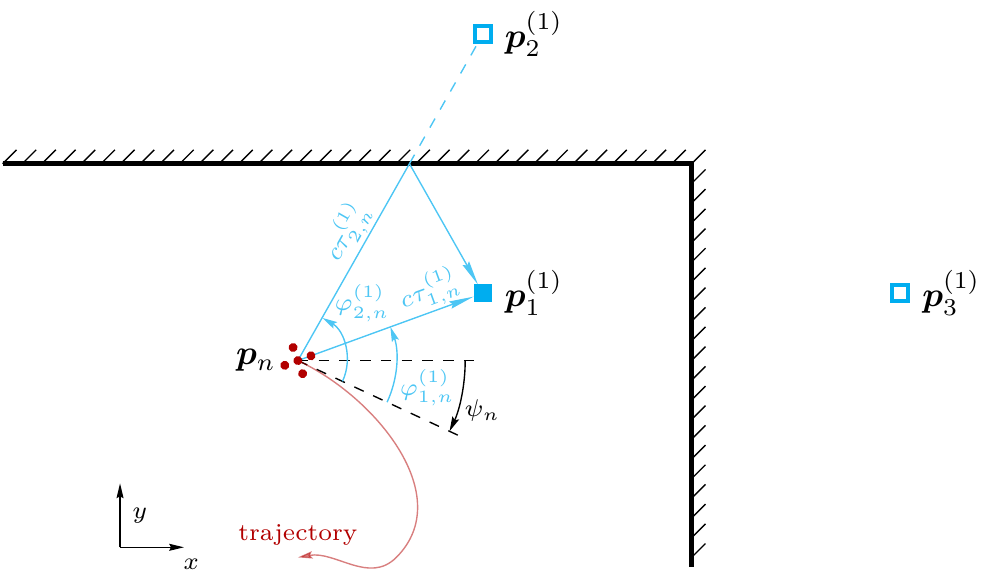}
\caption{Exemplary environment in a room corner. The mobile agent at unknown position $\V{p}_n$ is equipped with an array, indicated by red dots. The PA at position $\V{p}_1^{(1)}$ is marked by a cyan box. The cyan squares outside the room indicate some VAs associated with the PA.}\label{fig:fp} 
\end{figure}

New systems supporting multipath channels take advantage of it by exploiting multipath components (MPCs) for localization \cite{GentnerTWC2016, LeitingerTWC2019, WymeerschGLOBECOM2018, MendrzikJSTSP2019, LiLeitingerTWC2019}. A promising approach is to model specular reflections included in the radio signal at flat surfaces by virtual anchors (VAs) that are mirror images of the physical anchors (PAs) \cite{WitrisalSPM2016}. Multipath-based SLAM algorithms can detect and localize reflective flat surfaces represented by VAs, and jointly estimate the time-varying position of mobile agents. Fig.~\ref{fig:fp} shows an exemplary environment highlighting the geometric relations between the agent and physical anchors (PAs) or virtual anchors (VAs).

In this work, we extended a multipath-based SLAM belief propagation (BP) algorithm \cite{LeitingerTWC2019, LeitingerICC2019} by interacting multiple models (IMM) parameters to describe the agent state dynamics based on \cite{SoldiTSP2019,SoldiFUSION2018}. The time-evolution of the interacting multiple models (IMM) parameters is described by a Markov chain and the IMM parameters are incorporated into the factor graph structure that represents the statistical structure of the SLAM problem. 
The proposed belief propagation (BP)-based algorithm adapts in an online manner to time-varying IMM parameters by jointly inferring the model parameters along with the agent and map feature states. The algorithm jointly performs probabilistic data association (DA) and sequential Bayesian estimation of the state of a mobile agent and the states of ``potential VA'' (PVAs) characterizing the environment \cite{MeyerProc2018, MeyerICC2018, LeitingerTWC2019}. The states are augmented by a binary existence variable, associated to a \emph{probability of existence} enabling the estimation of an unknown number of VAs.

\section{Signal Model}\label{sec:signalmodel}
The agent is equipped with an $H$-element antenna array. At every discrete time step $n$, the element locations are denoted by $\V{p}_n^{(h)}$, $h \in \{1,\dots,H\}$, the agent position $\V{p}_n$ refers to the center of gravity of the array. We also define $d^{(h)} = \|\V{p}_n^{(h)} - \V{p}_n\|$ and $\psi^{(h)} = \angle{(\V{p}_n^{(h)}-\V{p}_n)}-\psi_n$, the distance from the reference location $\V{p}_n$ and the orientation, respectively, of the $h$-th element. The array orientation $\psi_n$ is unknown.

At every time $n$, anchor $j \in \{1,\dots, J\}$ transmits a baseband signal $s(t)$ centered at carrier frequency $f_\text{c}$. The signal at antenna element $h$ is then given as 
\begin{equation}\label{eq:rx_signal}
  s_{\textrm{RX},n}^{(j,h)}(t) = \sum_{l=1}^{L_n^{(j)}}\alpha_{l,n}^{(j,h)} \ist s\big(t \rmv-\rmv \tau_{l,n}^{(j,h)}\big) + \ist \nu_{n}^{(j,h)}(t) \ist+\ist w_n^{(j,h)}(t)\ist.
\end{equation}
The first term on the right-hand side (RHS) describes $L_n^{(j)}$ specular MPCs with complex amplitudes $\alpha_{l,n}^{(j,h)}=\alpha_{l,n}^{(j)}\exp\big(i 2 \pi f_\mathrm{c} \frac{d^{(h)}}{c} \cos\big(\varphi_{l,n}^{(j)}-\psi_n-\psi^{(h)}\big)\big)$ and delays $\tau_{l,n}^{(j,h)}=\tau_{l,n}^{(j)}-\frac{d^{(h)}}{c} \cos(\varphi_{l,n}^{(j)}-\psi_n-\psi^{(h)})$ with $\tau_{l,n}^{(j)} \rmv=\rmv \big\|\V{p}_{n} \!-\rmv \V{p}_{l}^{(j)}\big\|/c$ with $c$ as speed of light and AoA  $\varphi^{(j)}_{l,n} = \angle{(\V{p}_{l}^{(j)}-\V{p}_n)}-\psi_n$ related to the distance and angle between the agent's position $\V{p}_{n} \in \mathbb{R}^2$ and the PA/VA positions $\V{p}_{l}^{(j)} \in \mathbb{R}^2$, respectively. The second term on the RHS of \eqref{eq:rx_signal}, $\nu_{n}^{(j,h)}(t)$, represents the DM, which interferes with the position-related, specular MPCs. The last term on the RHS of \eqref{eq:rx_signal}, $w_n^{(j,h)}(t)$, is AWGN with power spectral density $N_0$. 

\textbf{Channel estimation}: A sparse Bayesian multipath channel estimator \cite{BadiuTSP2017, HansenTSP2018} is used to estimate at each time $n$ and for each PA a set of $M^{(j)}_n$ MPC distances $\hat{d}^{(j)}_{m,n}=c\hat{\tau}^{(j)}_{m,n}$ and AoAs $\hat{\varphi}^{(j)}_{m,n}$ and the corresponding complex amplitudes $\hat{\alpha}^{(j)}_{m,n}$, where $m \rmv\in\rmv \Set{M}_n^{(j)} = \{ 1,\dots,M_n^{(j)} \}$. To infer the variances $\hat{\sigma}^{(j)\ist2}_{\alpha,m,n}$ of the amplitudes, we use the antenna array at the agent. The measured normalized amplitude is given as $\hat{u}^{(j)}_{m,n} = |\hat{\alpha}^{(j)}_{m,n}|/\hat{\sigma}^{(j)}_{\alpha,m,n}$. The MPC parameters are combined into the measurement vector $\V{z}^{(j)}_{n} = [\V{z}^{(j) \ist \mathrm{T}}_{1,n},\dots,\V{z}^{(j) \ist \mathrm{T}}_{M^{(j)}_n,n}]\trans$, i.e., $\V{z}^{(j)}_{m,n}=[\hat{d}^{(j)}_{m,n}, \hat{\varphi}^{(j)}_{m,n}, \hat{u}^{(j)}_{m,n}]\trans$. The vectors $\V{z}_n^{(j)}$ are used as noisy ``measurements'' by the SLAM algorithm.

\section{System Model}\label{sec:sysmodel}
At each time $n$, the state $\V{x}_n $ of the agent consists of it's position, velocity, and orientation. As in \cite{MeyerProc2018, LeitingerTWC2019, LeitingerICC2019}, we account for the unknown number of features by introducing for each PA $j$ potential VAs (PVAs) $k \rmv\in\rmv \{ 1,\dots,K_n^{(j)} \}$. The number of PVAs $K_n^{(j)}$ is the maximum possible number of actual features that produced measurements so far \cite{MeyerProc2018} (where $K_n$ increases with time). The state of PVA $(j,k)$ are denoted as $\V{y}_{k,n}^{(j)} \!\triangleq\rmv \big[\V{x}_{k,n}^{(j)\text{T}}\,\ist r_{k,n}^{(j)}\big]\trans$ with vector $\V{x}_{k,n}^{(j)} = \big[\V{p}_{k}^{(j)\text{T}} \,\ist u_{k,n}^{(j)} \big]\trans$, which comprises the fixed PVA position $\V{p}_{k}^{(j)}$ and the normalized amplitude $u_{k,n}^{(j)}$ \cite{LeitingerICC2019}. The existence/nonexistence of PVA $k$ is modeled by the existence variable $r^{(j)}_{k,n} \rmv\in \{0,1\}$ in the sense that PVA $k$ exists if and only if $r^{(j)}_{k,n} \!=\! 1$. It is considered formally also if PVA $k$ is nonexistent, i.e., if $r^{(j)}_{k,n} \!=\rmv 0$. The states $\V{x}^{(j)\ist\text{T}}_{k,n}$ of nonexistent PVA are obviously irrelevant. Therefore, all probability density functions (PDFs) defined for PVA states $f(\V{y}^{(j)}_{k,n}) =\rmv f(\V{x}^{(j)}_{k,n}, r^{(j)}_{k,n})$ are of the form $f(\V{x}^{(j)}_{k,n}, 0 )$ $=\rmv f^{(j)}_{k,n} f_{\text{d}}(\V{x}^{(j)}_{k,n})$, where $f_{\text{d}}(\V{x}^{(j)}_{k,n})$ is an arbitrary ``dummy PDF'' and $f^{(j)}_{k,n} \!\rmv\in [0,1]$ is a constant. We also define the stacked vectors $\V{y}_n^{(j)} \!\triangleq \big[\V{y}_{1,n}^{(j)\text{T}} \rmv\cdots\ist \V{y}_{K_n^{(j)}\rmv,n}^{(j)\text{T}} \big]\trans$ and $\V{y}_n \!\triangleq \big[\V{y}_n^{(1)\text{T}} \rmv\cdots\ist \V{y}_n^{(J)\text{T}} \big]\trans\rmv$.

At any time $n$, each PVA is either a \textit{legacy PVA}, which was already established in the past, or a \textit{new PVA}, which is established for the first time. The states of legacy PVAs and new PVAs for PA $j$ will be denoted by $\underline{\V{y}}_{k,n}^{(j)} \!\triangleq\rmv \big[ \underline{\V{x}}_{k,n}^{(j)\text{T}} \; \underline{r}_{k,n}^{(j)} \big]\trans\rmv$ with $k \rmv\in\rmv \{ 1,\dots,K_{n-1}^{(j)} \}$ and $\overline{\V{y}}_{m,n}^{(j)} \!\triangleq\rmv \big[ \overline{\V{x}}_{m,n}^{(j)\text{T}} \; \overline{r}_{m,n}^{(j)} \big]\trans\rmv$ with $m \rmv\in\rmv \{ 1,\dots,M_n^{(j)} \}$, respectively. Thus, the number of new PVAs equals the number of measurements, $M^{(j)}_n \rmv$. The set and number of legacy PVAs are updated according to $K_{n}^{(j)} \rmv= K_{n-1}^{(j)} + M_{n}^{(j)}$. 

The vector of all legacy PVA states and new PVA states at time $n$ for PA $j$ are given by $\underline{\V{y}}_n^{(j)} \!\triangleq \big[\underline{\V{y}}_{1,n}^{(j)\text{T}} \rmv\cdots\ist \underline{\V{y}}_{K_{n-1}^{(j)}\rmv,n}^{(j)\text{T}} \big]\trans\rmv$ and $\overline{\V{y}}_n^{(j)} \!\triangleq \big[\overline{\V{y}}_{1,n}^{(j)\text{T}} \rmv\cdots\ist \overline{\V{y}}_{M_{n}^{(j)}\rmv,n}^{(j)\text{T}} \big]\trans\rmv$, respectively. For all PAs the states of all legacy and new PVAs are given by the vectors $\underline{\V{y}}_n \!\triangleq \big[\underline{\V{y}}_n^{(1)\text{T}} \rmv\cdots\ist \underline{\V{y}}_n^{(J)\text{T}} \big]\trans\rmv$ and $\overline{\V{y}}_n \!\triangleq \big[\overline{\V{y}}_n^{(1)\text{T}} \rmv\cdots\ist \overline{\V{y}}_n^{(J)\text{T}} \big]\trans\rmv$, respectively. The state of all PVAs at time $n$ is given by the vector $\V{y}_{n} \!\triangleq \big[\underline{\V{y}}_n^{\text{T}} \,\ist \overline{\V{y}}_n^{\text{T}}\big]\trans\rmv$.

\subsection{Association Vectors}
\label{sec:assoc_vec_description}

For each PA, measurements $\V{z}_{m,n}^{(j)}$, are subject to a DA uncertainty. It is not known which measurement $\V{z}_{m,n}^{(j)}$ is associated with which PVA $k$, or if a measurement $\V{z}_{m,n}^{(j)}$ did not originate from any PVA (\emph{false alarm}) or if a PVA did not give rise to any measurement (\emph{missed detection}). The probability that a PVA is ``detected'', in the sense that it generates a measurement $\V{z}_{m,n}^{(j)}$ in the MPC parameter estimation stage, is denoted by $P_\text{d}(\V{x}_n ,\V{x}^{(j)}_{k,n}) \rmv\rmv\triangleq\rmv\rmv P_{\textrm{d}}(u_{k,n}^{(j)} )$, being defined by its normalized amplitude $u_{k,n}^{(j)}$. The distribution of false alarm measurements $f_\textrm{FA}(\V{z}_{m,n}^{(j)})$ is assumed to be known. The associations between measurements $\V{z}_{m,n}^{(j)}$ and the PVAs at time $n$ can be described by the $K$-dimensional feature-oriented DA vector $\V{c}_n^{(j)} = [c_{1,n}^{(j)} \cdots\ist c_{K,n}^{(j)} ]\trans$, with entries $c_{k,n}^{(j)} =\rmv M \rmv\in\rmv \{ 1,\dots,M_n^{(j)} \}$, if PVA $k$ generates $\V{z}_{m,n}^{(j)}$ and $0$, if PVA $k$ does not generate any $\V{z}_{m,n}^{(j)}$. In addition, we consider the $M_n$-dimensional measurement-oriented DA vector $\V{b}_{n}^{(j)} = [b_{1,n}^{(j)} \cdots\ist b_{M_n,n}^{(j)}]\trans$ with entries $b_{m,n}^{(j)} =\rmv k \rmv\in\rmv \{ 1,\dots,K_{n-1}^{(j)} \}$, if $\V{z}_{m,n}^{(j)}$ is generated by PVA $k$ and $0$ if not generated by any PVA \cite{WilliamsTAE2014,WilliamsTAES2015,MeyerProc2018}.

\subsection{State Evolution}
\label{sec:state_statistics}

Following the interacting multiple model (IMM) approach \cite{BarShalom2002EstimationTracking}, the agent state can switch between different dynamic models $ q_{n} \rmv\rmv\in\rmv\rmv\mathcal{Q}\rmv\triangleq \rmv \{1,\cdots,Q\} $ at any time $ n $. The according state-transition PDF of the agent state is given by $ f_{q_{n}}(\bm{x}_{n}|\bm{x}_{n-1}) $. The dynamic mode (DM) index $ q_{n} $ is modeled as a random variable which evolve according to the first-order Markov chain with a constant transition matrix $ \bm{Q} \in [0,1]^{Q \times Q} $ over time ($ [0,1]^{Q \times Q} $ denotes a $ Q \rmv\rmv \times\rmv\rmv Q $ matrix with entries between $ 0 $ and $ 1 $). The DM's transition probability mass function (pmf) of $ q_{n} $ is given by $ p(q_{n}\rmv\rmv =\rmv\rmv i \rmv | \rmv q_{n-1}\rmv\rmv =\rmv\rmv i') = [\bm{Q}]_{i,i'} $ for $ i,i' \in \mathcal{Q} $. Note that $ \sum_{i' = 1}^{Q} [\bm{Q}]_{i,i'} = 1 $ $ \forall \, i $. The target state $ \bm{x}_{n} $ and the DM index $ q_{n} $ are assumed to jointly evolve according to a Markovian dynamic model, i.e., $f(\V{x}_{n},q_{n}|\V{x}_{n-1},q_{n-1}) = f_{q_{n}}(\V{x}_{n}|\V{x}_{n-1})f(q_{n}|q_{n-1})$.

The agent state $\V{x}_n$ and the augmented states of the legacy PVAs, $\underline{\V{y}}_{k,n}^{(j)}$, are assumed to evolve independently according to Markovian state dynamics, i.e., 
\begin{align}
&f\big(\V{x}_n,\underline{\V{y}}_n, q_{n}|\V{x}_{n-1},\V{y}_{n-1},q_{n-1}\big) \nn \\
&= f_{q_{n}}(\V{x}_{n}|\V{x}_{n-1})f(q_{n}|,q_{n-1})
\prod_{j=1}^J \prod_{k=1}^{K_{n-1}^{(j)}} \rmv\rmv f\big(\underline{\V{y}}_{k,n}^{(j)} \big| \V{y}_{k, n-1}^{(j)}\big)\\[-7mm]\nn
\end{align}
where $f(\underline{\V{y}}_{k,n}^{(j)} | \V{y}_{k, n-1}^{(j)})$ is the state-transition PDFs of legacy PVA $(j,k)$. Note that $\underline{\V{y}}_{k,n}^{(j)}$ depends on both $\underline{\V{y}}_{k,n-1}^{(j)}$ and $\overline{\V{y}}_{m,n-1}^{(j)}$. If PVA $(j,k)$ existed at time $n \rmv-\! 1$, i.e., $r_{k,n-1}^{(j)} \!=\! 1$, it either dies, i.e., $\underline{r}_{k,n}^{(j)} \!=\rmv 0$, or survives, i.e., $\underline{r}_{k,n}^{(j)} \!=\! 1$; in the latter case, it becomes/remains a legacy PVA at time $n$. The probability of survival is denoted by $P_\text{s}$. If the PVA survives, its new PMVA state $\underline{\V{y}}_{k,n}^{(j)}$ is distributed according to the state-transition PDF $f\big(\underline{\V{y}}_{k,n}^{(j)} \big| \underline{\V{y}}_{k,n-1}^{(j)}\big)$ (for details cf. \cite{LeitingerTWC2019}). It it assumed that at time $n \rmv=\rmv 1$ for each PA $j$ the initial prior PDFs $f\big(\V{y}^{(j)}_{k,1} \big)$, $k = \big\{1,\dots,K^{(j)}_1\big\}$ and $f(\V{x}_{1})$ are known. All (legacy and new) PVA states up to time $n$ are denoted as $\V{y}_{1:n} \triangleq \big[\V{y}^{\text{T}}_{1} \cdots\ist \V{y}^{\text{T}}_{n} \big]^{\text{T}}\!$.

\subsection{Joint Posterior PDF and the FG}
\label{sec:derivationFactorGraph}

By using common assumptions \cite{MeyerProc2018,LeitingerTWC2019}, and for fixed and thus observed measurements $\V{z}_{1:n}$, it can be shown that the joint posterior PDF of $\V{x}_{0:n}$ ($\V{x}_{1:n} \triangleq [\V{x}_1\trans \cdots \V{x}_n\trans]\trans$), $\V{q}_{1:n}$, $ \V{c}_{1:n}$, $\V{y}_{0:n}$, and $\V{b}_{1:n}$, conditioned on $\V{z}_{1:n}$ for all time steps is given by
\begin{align}
&f( \V{x}_{1:n}, \V{q}_{1:n}, \V{y}_{1:n}, \V{c}_{1:n}, \V{b}_{1:n}| \V{z}_{1:n} ) \nn\\[-1.1mm]
&\hspace{1mm}\propto  f(\V{x}_{1}) \Bigg(\prod^{J}_{j'=1} \rmv 
  \prod^{K^{(j')}_{1}}_{k'=1} \! f\big( \V{y}^{(j')}_{k'\!,1}\big) \Bigg)  \nn\\[-0.9mm]
&\hspace{2mm}\times\rmv\ \prod^{n}_{n'=2}  \! f_{q_{n'}}(\V{x}_{n'}|\V{x}_{n'-1})f(q_{n'}|,q_{n'-1}) \nn\\[-0.9mm]
&\hspace{2mm}\times\rmv\ \prod^{J}_{j=1}  \Bigg(\prod^{K^{(j)}_{n'-1}}_{k=1} f\big(\underline{\V{y}}^{(j)}_{k,n'} \big| \V{y}^{(j)}_{k,n'-1}\big)\nn\\[-1.1mm]
&\hspace{2mm}\times  \rmv g\big( \V{x}_{n'}, \underline{\V{y}}^{(j)}_{k,n'}, c^{(j)}_{k,n'}; \V{z}^{(j)}_{n'} \big)\rmv \prod^{M^{(j)}_{n'}}_{m'=1}\rmv \Psi_{km'}\big(c^{(j)}_{k,n'} \rmv,b^{(j)}_{m',n'}\big) \Bigg) \nn\\[-1mm]
&\hspace{2mm}\times \prod^{M^{(j)}_{n'}}_{m=1} f(\overline{\V{y}}^{(j)}_{m,n}) h\big( \V{x}_{n'}, \overline{\V{y}}^{(j)}_{m,n'}, b^{(j)}_{m,n'}; \V{z}^{(j)}_{n'} \big)
\label{eq:factorization_post} 
\end{align}
where $g\big( \V{x}_{n}, \underline{\V{y}}^{(j)}_{k,n}, c^{(j)}_{k,n}; \V{z}^{(j)}_{n} \big))$,  $f(\overline{\V{y}}_{m,n})$,  $\Psi_{km}(a^{(j)}_{k,n},b^{(j)}_{m,n})$, and $ h\big(\V{x}_{n}, \overline{\V{y}}^{(j)}_{m,n}, b^{(j)}_{m,n}; \V{z}^{(j)}_{n'} \big)$ are explained in what follows.

The pseudo likelihood functions $g\big( \V{x}_{n}, \underline{\V{y}}^{(j)}_{k,n}, c^{(j)}_{k,n}; \V{z}^{(j)}_{n} \big)) = g\big( \V{x}_{n}, \underline{\V{x}}^{(j)}_{k,n}, \underline{r}^{(j)}_{k,n}, c^{(j)}_{k,n}; \V{z}^{(j)}_{n} \big))$ and $ h\big(\V{x}_{n}, \overline{\V{y}}^{(j)}_{m,n}, b^{(j)}_{m,n}; \V{z}^{(j)}_{n'} \big) = h\big(\V{x}_{n}, \overline{\V{x}}^{(j)}_{m,n}, \overline{r}^{(j)}_{m,n}, b^{(j)}_{m,n}; \V{z}^{(j)}_{n'} \big)$ are given by
\begin{align}
&g\big( \V{x}_{n},  \underline{x}^{(j)}_{k,n}, 1, c^{(j)}_{k,n}; \V{z}^{(j)}_{n} \big) \nn \\
&\hspace{4mm}\triangleq \begin{cases}
  \rmv\displaystyle \frac{P_{\mathrm{d}}\big(u^{(j)}_{m,n}\big) f\big(\V{z}^{(j)}_{m,n} \big|\V{x}_n, \underline{\V{x}}^{(j)}_{k,n} \big)}{\mu_{\mathrm{fa}}f_{\text{fa}}\big( \V{z}^{(j)}_{m,n} \big)} \ist, 
  &\!\rmv\rmv\rmv c^{(j)}_{k,n} \!=\rmv m  \\[-.5mm]
  \rmv 1 \ist, & \!\rmv\rmv\rmv c^{(j)}_{k,n} \!=\rmv 0
\end{cases}
\end{align}
and $g\big( \V{x}_n, \underline{\V{x}}^{(j)}_{k,n} , 0, c^{(j)}_{k,n}; \V{z}^{(j)}_{n} \big) \rmv\triangleq\rmv \delta \big(c^{(j)}_{k,n}\big)$ as well as 
\begin{align}
&h\big( \V{x}_n, \overline{\V{x}}^{(j)}_{m,n} , 1, b^{(j)}_{m,n}; \V{z}^{(j)}_{n} \big) \nn \\
&\hspace{4mm}\triangleq \begin{cases}
     0 \ist,  & \! b^{(j)}_{m,n} \!\in\rmv\Set{K}^{(j)}_n \\[.5mm]
     \displaystyle \frac{f\big( \V{z}^{(j)}_{m,n} \big| \V{x}_n, \overline{\V{x}}^{(j)}_{m,n} \big) }{ \mu_{\mathrm{fa}} f_{\text{fa}}\big( \V{z}^{(j)}_{m,n} \big)} 
  \ist, &\! b^{(j)}_{m,n} \rmv=\rmv 0
  \end{cases}\label{eq:factor_newPVAs}
\end{align}
and $h\big( \V{x}_n, \overline{\V{x}}^{(j)}_{m,n} , 0, b^{(j)}_{m,n}; \V{z}^{(j)}_{n} \big) \rmv\triangleq\rmv f_{\text{D}}\big(\overline{\V{x}}^{(j)}_{m,n}\big)$. 

The \textit{prior distributions} $f(\overline{\V{y}}_{m,n}) \rmv=\rmv f(\overline{\V{x}}^{(j)}_{m,n}, \overline{r}_{m,n})$ for new PVA state $k \rmv\in\rmv \{1,\dots,M^{(j)}_n\}$ can be expressed as
\begin{align}
f(\overline{\V{x}}^{(j)}_{m,n}, \overline{r}_{m,n})  &\,\triangleq \begin{cases}
      \mu_{\mathrm{n}} \ist f_{\mathrm{n}}(\overline{\V{x}}^{(j)}_{m,n})  \ist, 
       & \rmv\rmv \overline{r}_{m,n}= 1 \\[1mm]
     f_{\mathrm{d}}\big(\overline{\V{x}}^{(j)}_{m,n}\big) \ist,  & \rmv\rmv \overline{r}_{m,n} = 0 \ist.
  \end{cases}
\end{align}
Finally, the binary \textit{indicator functions} that check consistency for any pair of PVA-oriented  and measurement-oriented association variable at time $n$ read
\begin{align}
&\Psi_{km}(a^{(j)}_{k,n},b^{(j)}_{m,n}) \nn\\[1.3mm]
&\hspace{2mm}\triangleq \begin{cases} 
   0, & a^{(j)}_{k,n} =  m, b^{(j)}_{m,n} \neq k \text{ or } a^{(j)}_{k,n} \neq  m, b^{(j)}_{m,n} = k  \\[.1mm]
   1 \ist, & \text{otherwise}. 
  \end{cases}
\end{align}
In case the joint PVA-oriented association vector $\V{p}_n$ and the measurement-oriented association vector $\V{b}_n$ do not describe the same association event, at least one indicator function in \eqref{eq:factorization_post}  is zero and thus $f( \V{y}_{0:n}, \V{x}_{0:n}, \V{p}_{1:n},\V{b}_{1:n} \ist | \ist \V{z}_{1:n})$ is zero as well. The factor graph \cite{KschischangTIT2001} representing factorization \eqref{eq:factorization_post} is very similar to the ones in \cite{LeitingerTWC2019} or \cite{LeitingerICC2019}. 

\subsection{Minimum Mean-Square Error (MMSE)  Estimation}\label{sec:problem}

Our goal is to estimate the agent state $\V{x}_{n}$ and the positions $\V{p}_{k}$ and the amplitudes $u_{k,n} $ of the PVAs from past and present measurements, i.e., from the total measurement vector $\V{z}_{1:n}$. In the Bayesian framework, estimation of the states are based on their respective posterior PDFs. We develop an approximate calculation of the minimum mean-square error (MMSE) estimates of the agent state $\V{x}_n$, the positions $\V{p}_{k}$ and the amplitudes $u_{k,n} $ of the PVAs based on the marginal posterior PDFs:
\begin{align}
\hat{\V{x}}^\text{MMSE}_{n} \,&\triangleq\rmv \int\rmv \V{x}_n \ist f(\V{x}_n |\V{z}_{1:n}) \ist \mathrm{d}\V{x}_n
\label{eq:mmse_agent} \\
\hat{q}^\text{MMSE}_{n} \,&\triangleq\rmv \sum_{i \in \mathcal{Q}}\ist i \ist f(q_n=i |\V{z}_{1:n})
\label{eq:mmse_IMM} \\
\hat{\V{x}}^{(j)\ist\text{MMSE}}_{k,n} \,&\triangleq\rmv \int\rmv \V{x}^{(j)}_{k,n}\, f\big(\V{x}^{(j)}_{k,n} \big| r^{(j)}_{k,n}=1, \V{z}_{1:n}\big) \ist \mathrm{d}\V{x}^{(j)}_{k,n} \label{eq:MMSE_PVApos} 
\end{align}
where $f(\V{x}_n |\V{z}_{1:n}) = \sum_{q_n \in \mathcal{Q}}f(\V{x}_n, q_n|\V{z}_{1:n})$ with the joint marginal PDF of the agent state and the dynamic model state $f(\V{x}_n, q_n|\V{z}_{1:n})$ \cite{SoldiFUSION2018}, 
\begin{equation}
f\big(\V{x}^{(j)}_{k,n} \big| r^{(j)}_{k,n}=1, \V{z}_{1:n}\big)= \frac{f\big(\V{x}^{(j)}_{k,n}, r^{(j)}_{k,n}=1 \big| \V{z}_{1:n}\big)}{p\big(r^{(j)}_{k,n}=1 \big| \V{z}_{1:n}\big)}\ist,\\[0mm]
\end{equation}
and $p\big(r^{(j)}_{k,n}=1 \big| \V{z}_{1:n}\big)=\int\rmv f\big(\V{x}^{(j)}_{k,n}, r^{(j)}_{k,n}=1 \big| \V{z}_{1:n}\big) \ist \mathrm{d}\V{x}^{(j)}_{k,n}$. The state of the $k$-th PVA is only estimated if it is considered as detected at time $n$, i.e., $p\big(r^{(j)}_{k,n}=1 \big| \V{z}_{1:n}\big)>P_\mathrm{det}$ with detection probability threshold $P_\mathrm{det}$.

By performing sequential particle-based message passing by means of the SPA rules \cite{KschischangTIT2001,MeyerTSP2017,LeitingerTWC2019} on the factor graph in \cite{LeitingerTWC2019}, approximations (``beliefs'') of the marginal posterior PDFs $f(\V{x}_n, q_n|\V{z}_{1:n})$, $p(r^{(j)}_{k,n} \!=\! 1 |\V{z})$, and $f\big(\V{x}^{(j)}_{k,n}, r^{(j)}_{k,n}=1 \big| \V{z}_{1:n}\big)$ can be obtained in an efficient way for the agent state as well as all legacy and new PVA states for each PA. Details about the messages can be found in \cite{LeitingerTWC2019} and \cite{MeyerTSP2017,SoldiTSP2019}. To avoid that the number of PVA states grows to infinity, PVA states with a probability of existence $p\big(r^{(j)}_{k,n}=1 \big| \V{z}_{1:n}\big)$ below a threshold $p_{\text{pr}}$ are removed from the state space (``pruned'') after processing the measurements of each PA.

\section{Performance Evaluation}
\begin{figure}
\centering
\includegraphics[width=0.8\columnwidth]{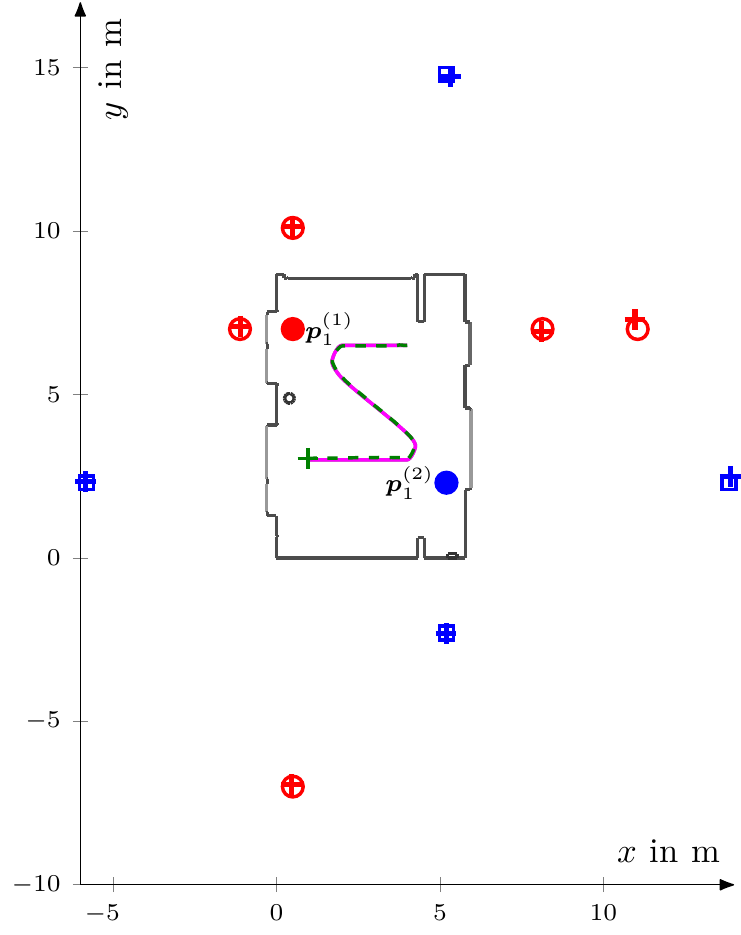}
\caption{Floorplan of the emulated room including the agent track (solid magenta), the position of PA 1 (red filled circle) and PA 2 (blue filled circle),  the according VAs (red and blue circles for PA 1 and PA 2, respectively), the estimated agent track (green dashed), and the estimated VAs (red and blue pluses for PA 1 and PA 2, respectively).}\label{fig:fp_track} 
\end{figure}

\begin{figure}[t]
\captionsetup[subfigure]{captionskip=0pt}
\subfloat[\label{fig:OSPA_Agent}]{\includegraphics[width=1\columnwidth,height=0.35\columnwidth]{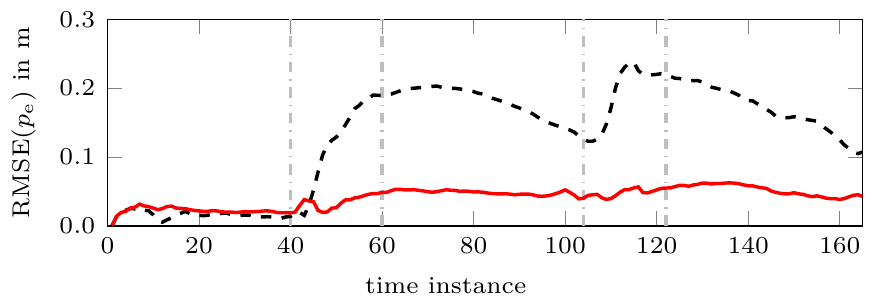}}\\[-0.5mm]
\subfloat[\label{fig:OSPA_MAP}]{\includegraphics[width=1\columnwidth,height=0.35\columnwidth]{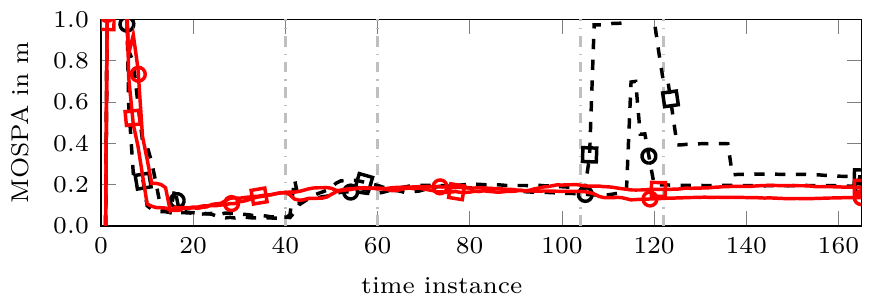}}\\[1.5mm]
\centering%
\includegraphics[width=1\columnwidth]{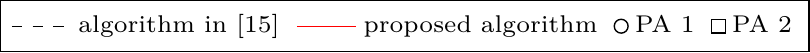}
\\[-1mm]
\caption{MOSPA error on the PVA positions (a) for PA 1 and PA 2. RMSE of the agent position (b).} 
\vspace*{-2mm}
\end{figure}

\begin{figure}[t]
\includegraphics[width=1\columnwidth,height=0.35\columnwidth]{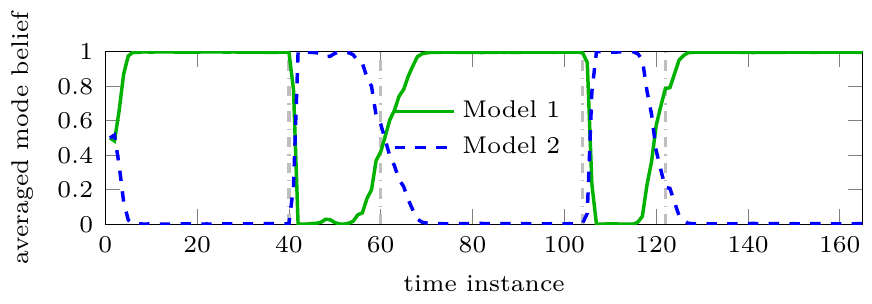}%
\caption{Average mode belief.} 
\label{fig:averageBelief}
\end{figure}

The proposed SLAM algorithm is validated in the indoor scenario shown in Fig.~\ref{fig:fp_track}. To show the benefit of the proposed algorithm, we have chosen an agent track that consists of straight parts with a small spacing between consecutive agent positions and two rapid turns with increased spacing between consecutive agent positions. The abrupt changes  of the agent state in direction are indicated by vertical gray lines in Fig.~\ref{fig:OSPA_Agent}, Fig.~\ref{fig:OSPA_MAP}, and Fig.~\ref{fig:averageBelief}. We compare the proposed SLAM algorithm with IMM to the original SLAM algorithm in \cite{LeitingerICC2019}.

The agent's state-transition pdfs $f_j(\V{x}_n|\V{x}_{n-1})$, with $\V{x}_n \rmv=\rmv  [\V{p}_n^{\mathrm{T}} \; \V{v}_n^{\mathrm{T}} ]^{\mathrm{T}}\rmv$, are defined by a near constant-velocity motion models \cite[Sec. 6.3.2]{BarShalom2002EstimationTracking}, where the driving process is iid across $n$, zero-mean, and Gaussian with driving process variance $\sigma_{w,q}^2$. The agent state can switch between two dynamic models ($Q =2$) with $\sigma_{w,1}^2 = 0.0032^2$ and $\sigma_{w,2}^2 = 0.01^2$. The dynamic model transition probabilities are chosen as $[\M{Q}]_{1,1} = [\M{Q}]_{2,2} = 0.99$ and $[\M{Q}]_{1,2} = [\M{Q}]_{2,1} = 0.01$. The original multipath-based SLAM from \cite{LeitingerICC2019} uses also a near constant-velocity motion model with constant driving process variance $\sigma_{w,1}^2 = 0.0032^2$.

For the sake of numerical stability, we introduced a small regularization noise to the PVA state $\V{p}^{(j)}_{k}$ at each time $n$, i.e., $\underline{\V{p}}^{(j)}_{k} \rmv\rmv=\rmv\rmv \V{p}^{(j)}_{k} \rmv+\rmv \V{\omega}_{k}$, where $\V{\omega}_{k}$ is iid across $k$, zero-mean, and Gaussian with covariance matrix $\sigma_a^2\bold{I}_2$ and $\sigma_a = 3\cdot10^{-3}\,\text{m}$.

We performed 10 simulation runs, each using the floor plan and agent track shown in Fig.~\ref{fig:fp_track}. The true MPC parameters, distance, AoA, and amplitude are calculated for the set of PAs and VAs as in Fig.~\ref{fig:fp_track}, with $5$ and $4$ VAs for PA $1$ and PA $2$, respectively.  All parameters related to the radio signal, the normalized amplitude, and the according detection probability can be found in \cite{LeitingerICC2019}. In each simulation run, we generated with detection probability $P_{\textrm{d}}(u^{(j)}_{m,n} )$ the noisy range measurements $\hat{d}^{(j)}_{m,n}$, AoA measurements $\hat{\varphi}^{(j)}_{m,n}$, and normalized amplitude measurements $\hat{u}^{(j)}_{m,n}$ related to the measurement likelihood functions in \cite[(7), (8), and (9)]{LeitingerICC2019}, respectively, using the true MPC parameters. In addition to the true noisy measurements we generate a mean number of $\mu_\mathrm{FA}^{(j)}=1$ false alarm measurements after thresholding for both PA $1$ and PA $2$ according to a false alarm pdf $f_{\mathrm{fa}}(z_{m,n}^{(j)})$ that is uniform on $[0\,\text{m},30\,\text{m}]$. The particles for the initial agent state are drawn from a 4-D uniform distribution with center $\V{x}_1 = [\V{p}_{0}^{\mathrm{T}}\;0\;\, 0]^{\mathrm{T}}\rmv$, where $\V{p}_{0}$ is the starting position of the actual agent track, and the support of each position component about the respective center is given by $[-0.1\,\text{m}, 0.1\,\text{m}]$ and of each velocity component is given by $[-0.05\,\text{m/s}, 0.05\,\text{m/s}]$.  The prior distribution for new PVA states $f_\mathrm{n}(\overline{\V{y}}_{m,n})$ is uniform on the square region given by $[-15\,\text{m},\ist 15\,\text{m}]\ist\times\ist[-15\,\text{m},\ist 15\,\text{m}]$ around the center of the floor plan shown in Fig.~\ref{fig:fp_track}. and the mean number of new PVs at time $n$ is $\mu_n = 0.05$. The probability of survival is $p_{\mathrm{s}} = 0.999$ and the detection and pruning thresholds are $p_{\mathrm{de}} = 0.5$ and $p_{\mathrm{pr}} = 10^{-3}$, respectively.

As an example, Fig.~\ref{fig:fp_track} depicts the MMSE estimates of the PVA positions for one simulation run as well as estimated agent tracks. 

Fig.~\ref{fig:OSPA_Agent} shows the root mean square error (RMSE) of the agent positions versus time $n$. The RMSE of the agent positions of the proposed algorithm stays below $7\ist$cm along the entire agent track, whereas the RSME of the agent position of algorithm presented in \cite{LeitingerICC2019} increases significantly after the first turn. Fig.~\ref{fig:OSPA_MAP} shows the mean optimal subpattern assignment (MOSPA) errors \cite{Schuhmacher2008} for the two PAs and the associated VAs, all versus time $n$. The MOSPA errors are based on the Euclidean metric with cutoff parameter $c = 1\,$m and order $p = 1$.  The solid lines show the MOSPA errors of the proposed SLAM algorithm and the dashed lines show the MOSPA errors of the algorithm presented in \cite{LeitingerICC2019}. The MOSPA error of the proposed SLAM algorithm and of the algorithm presented in \cite{LeitingerICC2019} are very similar until the second sharp turn starting at time $n=105$. After this time $n$, the MOSPA error of the algorithm presented in \cite{LeitingerICC2019} increases significantly in comparison to the proposed algorithm. However, in the majority of simulation runs, the original multipath-based SLAM algorithm presented in \cite{LeitingerICC2019} is also able to recover from this difficult situation.

Finally, Fig.~\ref{fig:averageBelief} shows the beliefs for the two dynamic models calculated by the proposed algorithm, averaged over the simulation runs. It can be seen that along the straight parts of the agent track with a smaller spacing between consecutive agent positions, dynamic model $1$ is in force, and whenever the change of direction and the spacing between consecutive agent positions is larger, dynamic model $2$ is in force.

\section{Conclusions}\label{sec:concl}
We have extended a multipath-based SLAM algorithm that continuously adapts IMM parameters to describe the mobile agent state dynamics. The time-evolution of the IMM parameters is described by a Markov chain and the parameters are incorporated into the factor graph. Our numerical simulations have shown that the proposed multipath-based algorithm is able to adapt the agent state space dynamic in an online manner enabling small agent errors even for rapid changes in the agent movement. 


\bibliographystyle{IEEEtran}
\bibliography{IEEEabrv,references}
\end{document}